# Designing an Integrated Learning Ecosystem for Introductory Python Programming: Scaffolded Problem Solving, Tablet-Inspired Navigation, and AI-Enhanced Instruction

**Xinxing Wu, Johnathan Rogers**

School of Mathematics and Computer Science, College of Business, Engineering and Technology
Kentucky State University, Frankfort, Kentucky, USA

Correspondence: xinxing.wu@kysu.edu

**Abstract:** Introductory programming requires students to represent problems, develop and test programs, debug errors, and apply computational ideas in new contexts. In this paper, we present an artifact-centered design analysis of an integrated learning ecosystem developed for a 16-lab introductory Python course. The analyzed corpus comprises 18 instructional artifacts: the course syllabus, a tablet-inspired homepage, and Labs 1-16. We examined course navigation, problem representation, scaffolding, independent transfer, application development, debugging, AI integration, and assessment. All 15 scaffolded labs shared seven elements: input-processing-output (IPO) reasoning, instructor-guided programming, an independent task, a Google Colab/Gradio extension, debugging support, a task checklist, and submission requirements. Lab 16 departed from this structure as an integrative Pygame-based Hallmark Assignment. The analysis identifies four connected design features: a navigation layer that coordinates multiple tools; two forms of transfer, from guided to independent problems and from console programs to interactive interfaces; distinct roles for instructor-side AI-enhanced communication and student-side AI assistance; and a transition from repeated scaffolds to integrative problem solving. The case also identifies unresolved tensions involving tool fragmentation, accessibility, interface distraction, student authorship, and overreliance on AI. Because student learning outcomes were not examined, we make no claims about instructional effectiveness. Instead, the paper offers a transparent design precedent and identifies priorities for future empirical evaluation.

**Demo** - https://xinxingwu-uk.github.io/COS107_Demo/Homepage/index.html

**Source codes -** https://github.com/xinxingwu-uk/Course_Integrated_Learning_Ecosystem

## 1. Introduction

Learning to program requires more than memorizing syntax. Beginning programmers must interpret a problem, identify inputs and expected outputs, select computational structures, develop executable solutions, and test and revise their code. These activities make programming both a technical skill and a form of problem solving [1-2]. Computational thinking offers a broader foundation by emphasizing problem formulation, representation, abstraction, automation, and evaluation rather than code production alone [3-4].

For novices, the difficulty is that these activities must often be coordinated at the same time. Students may reproduce a correct statement without knowing when to use it, how it contributes to a larger solution, or what to do when a program behaves unexpectedly. Prior work shows that beginning programmers may struggle to monitor their understanding and recognize an unproductive strategy [5], and similar conclusion argues that problem solving is often left implicit in programming exercises, even though novices can benefit from explicit metacognitive support [6].

Several computing-education frameworks address the transition from supported activity to independent creation. Use-Modify-Create asks students to engage with existing computational artifacts, modify them, and eventually create new ones [7]. One other study shows how this approach can balance structure with student agency [8].

PRIMM organizes learning around Predict, Run, Investigate, Modify, and Make, placing code reading and reasoning before independent production [9]. Although these frameworks differ, they share an important principle: novice programmers need visible support and a meaningful path toward independent problem solving.

A second challenge is connecting introductory code to authentic applications. Console programs are useful because they minimize interface complexity. However, students often encounter software through interactive interfaces. If early programming remains limited to console output, learners may struggle to see how basic constructs contribute to usable applications. Lightweight browser-based frameworks can connect foundational logic to interactive software without requiring a full web-development stack.

Generative artificial intelligence (GenAI) adds a third challenge. GenAI systems can explain syntax, suggest algorithms, identify errors, and generate complete programs from natural-language prompts. Researchers discuss the resulting opportunities and disruptions for computing education [10]. One recent study shows that such assistance can help novices who are able to evaluate it, while increasing difficulty for learners who cannot reliably judge generated solutions [11]. Producing a working program is therefore not equivalent to understanding it. AI-assisted programming raises a central instructional question: who performed the reasoning, and can the student explain and verify the result?

GenAI also creates opportunities for instructors to communicate course content through synthetic speech, music, video, and virtual avatars. Yet multimedia is not automatically educationally beneficial. Its value depends on coherence, signaling, learner characteristics, and instructional purpose [12]. Recent design cases likewise emphasize integrating GenAI in response to a clear learning need rather than for novelty alone [13-15].

This paper examines these challenges through COS 107: Problem Solving, Logic & Design, a three-credit introductory course at Kentucky State University. Python is used to develop computational thinking and program design. The course integrates Canvas, a tablet-inspired homepage, HTML-based instructional materials, Thonny, Google Colab, Gradio, GitHub resources, Matplotlib, SQLite, Pygame, and AI-enhanced course communication. The central design challenge was not how to introduce more tools, but how to organize existing tools into a coherent learning environment while shifting responsibility for computational reasoning from instructor to student.

The design combines several complementary elements. The tablet-inspired homepage serves as a course-level gateway. Labs 1-15 follow a recurring sequence of problem representation, guided coding, debugging, independent transfer, and interactive application development. Colab and Gradio connect console programs to user-facing interfaces. Instructor-side AI-generated media provide additional communication formats, while student AI assistance is subject to authorization and disclosure rules. Lab 16 departs from the regular sequence and serves as a Pygame-based Hallmark Assignment aligned with institutional problem-solving assessment.

We present the course as a design case supported by structured artifact analysis. Our aim is to document the artifacts, decisions, recurring patterns, tradeoffs, and unresolved questions so that other instructors and designers can examine the precedent. This study does not test whether the ecosystem improves learning, engagement, or retention. Instead, it asks four research questions about the design itself:

- **RQ1.** What recurring design structures characterize the integrated COS 107 learning ecosystem?
- **RQ2.** How does the laboratory sequence operationalize a progression from instructor-supported programming toward independent transfer and integrative problem solving?
- **RQ3.** How are tablet-inspired navigation and AI-enhanced instruction integrated as complementary layers without displacing foundational programming activity?
- **RQ4.** What design tensions, limitations, and future evaluation priorities emerge from the completed ecosystem?

## 2. Conceptual Framing

### 2.1 Computational thinking and problem representation

In COS 107, programming constructs are taught as resources for solving problems. Variables represent changing information; decisions encode alternatives; loops express repetition; functions support decomposition and reuse; collections organize related data; files and databases support persistence; visualization makes data interpretable; and classes coordinate state and behavior. The emphasis is on selecting structures to meet a problem's requirements, consistent with the problem-formulation orientation of computational thinking [3].

A recurring representation is the input-processing-output (IPO) model. Students identify the information a program needs, the operations it must perform, and the results it should produce. IPO provides a simple bridge between an informal problem statement and the computational structures used to implement a solution.

### 2.2 Scaffolding, modification, and learner agency

Instructional scaffolding helps students perform tasks that initially exceed their independent capabilities, with support adjusted as their competence develops. In programming, support must address both technical knowledge and the process of monitoring and revising a solution. COS 107 therefore uses a recurring sequence: concept introduction, problem representation, instructor-guided implementation, debugging, independent transfer, application extension, testing, and submission.

This sequence is related to Use-Modify-Create and PRIMM, but it does not reproduce either model exactly. Guided activities make a computational structure visible; independent tasks require students to adapt that structure to a related problem; and interface extensions ask them to reorganize the logic for a different form of interaction. The intended progression is toward greater student responsibility, not permanent reliance on templates.

### 2.3 Debugging as metacognitive practice

Debugging is presented as a normal part of programming rather than a separate activity undertaken only after failure. Early prompts ask students to locate an error, consider whether it involves spelling, punctuation, indentation, or a data type, and compare expected with observed behavior. Later labs introduce topic-specific common errors and multiple test cases. This progression moves attention from correcting syntax toward evaluating whether a program behaves as intended.

### 2.4 Authenticity through interactive applications

The course introduces program logic before interface construction. Students first work with Python in Thonny; after the core activity is established, they use Colab and Gradio to create an interactive version. A function or algorithm can therefore be examined as source code and then reused in a browser-based application. Gradio is not the main learning objective. Its role is to demonstrate how introductory Python can support usable software.

### 2.5 A functional view of AI in education

The ecosystem distinguishes AI use by instructional function. Instructor-side AI can present selected course information through multimodal materials such as singing-syllabus videos or talking avatars. Student-side AI may assist with brainstorming, explanations, debugging, data work, or code generation. The two roles have different implications. Instructor-side media require attention to clarity, accessibility, disclosure, and distraction; student-side assistance additionally raises questions about authorship, verification, dependency, and responsibility for the solution.

This distinction is consistent with human-centered guidance for GenAI in education [16] and work on aligning AI permissions with the intellectual work an assignment is intended to assess [13]. It also responds to evidence that novice programmers may not be able to evaluate AI-generated suggestions reliably [11].

## 3. Course Context and Design Problem

The institutional description of COS 107 emphasizes problem analysis, algorithm and pseudocode development, and data and control structures in a language-independent introduction to programming. Python is the language used in the current course. Learning outcomes include analyzing computing problems, designing algorithmic solutions using pseudocode, flowcharts, or IPO charts, translating those solutions into executable programs, and identifying and correcting code deficiencies.

Programming practice is central to assessment. Sixteen labs contribute 64 of the 100 regular course points; syllabus acknowledgement, attendance, a midterm, and a final examination account for the remainder. Fifteen optional challenges provide extension opportunities. The sequence progresses from output, strings, variables, and calculations to decisions, loops, functions, modules, collections, files, CSV data, visualization, object-oriented programming, SQLite, and an integrative project.

Each technology serves a different purpose. Canvas manages the institutional workflow; the custom homepage organizes course access; Thonny supports local beginner programming; Colab reduces installation demands; Gradio creates lightweight interactive interfaces; Matplotlib supports visualization; SQLite introduces structured data storage; and Pygame supports the culminating graphical program. These tools extend what students can build, but using several environments also creates a risk of fragmentation.

The tablet-inspired homepage was developed as a response to that risk. A browser-like area embeds the syllabus; a music-style area contains a course-structure video; a mail-style area displays the course-point structure; and a settings-style list provides access to the Introduction and Weeks 1-16. The homepage also links to Canvas. It functions as a course gateway, not a replacement for the institutional learning management system.

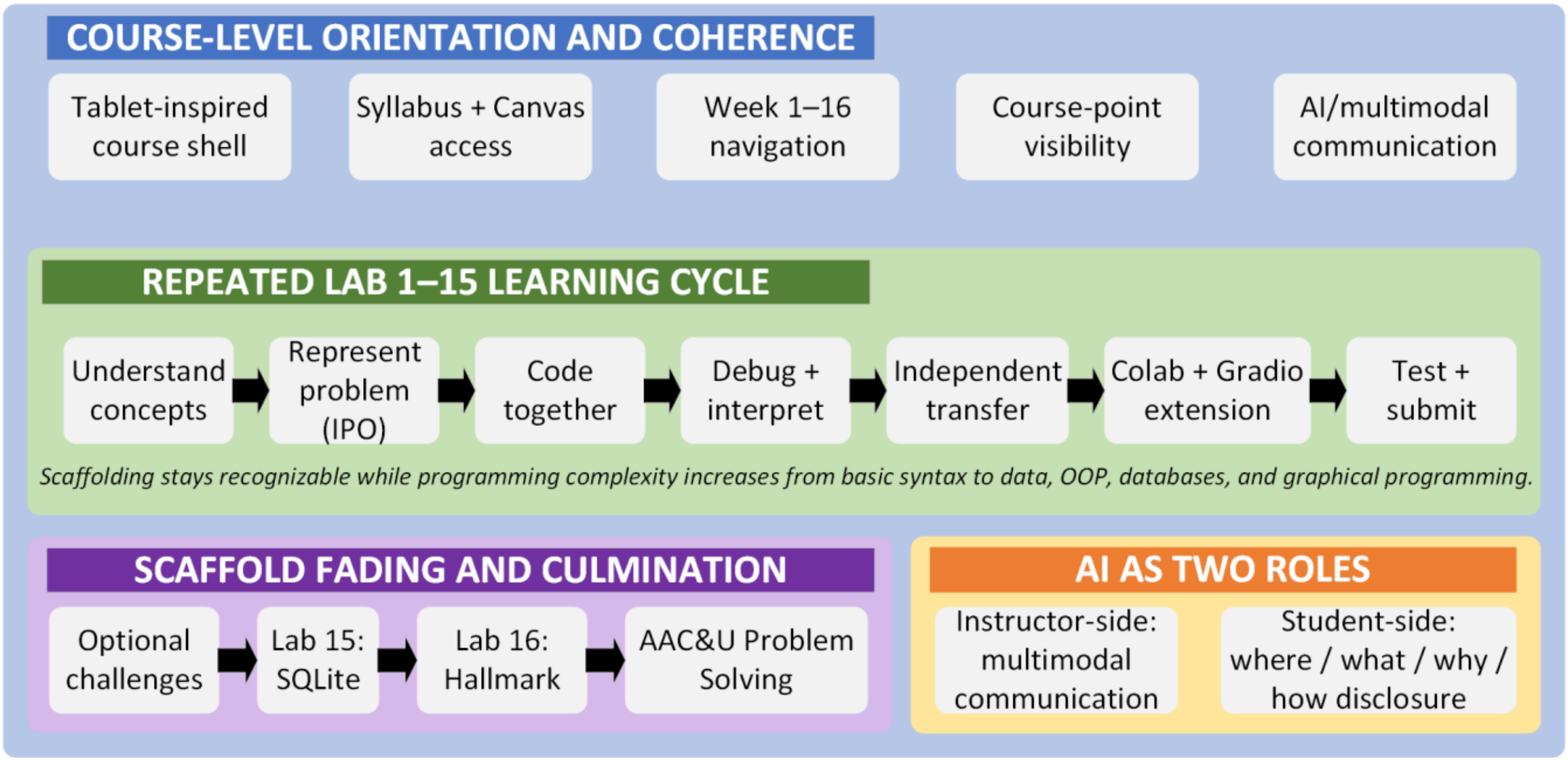

***Figure 1. Conceptual model of the COS 107 integrated learning ecosystem.***

AI introduces a parallel design challenge. The course uses instructor-side multimedia informed by prior work on singing syllabi and talking slide avatars [17-18]. Student-side AI is more restricted. The syllabus prohibits using AI to complete assignments unless its use is explicitly authorized, and any permitted use must be disclosed. Students must state where AI was used, identify the tool or model, explain why it was used, and describe how its output was incorporated. The syllabus also emphasizes responsible, legal, and ethical technology use.

The course culminates in a larger programming task. The syllabus identifies Lab 16 as the Hallmark Assignment for Essential Skill #2, 'Think critically to solve problems and create new ideas and solutions,' and specifies the AAC&U Problem Solving VALUE Rubric. The reviewed Lab 16 page contains a Pygame-based Flappy Bird codebase using modules, constants, classes, methods, randomization, events, keyboard input, update loops, sprite groups, image and audio assets, movement, and collision detection.

## 4. Method: Artifact-Centered Design Analysis

### 4.1 Design-case orientation

This study combines a design-case orientation with structured artifact analysis. Design cases document intentionally developed artifacts or experiences in enough detail to expose the decisions, constraints, tradeoffs, and unresolved questions that shaped them. Our unit of analysis is the course environment, not a sample of students.

We make claims about observable design features, not causal learning effects. For example, the artifacts can show that debugging prompts recur across the labs and that the homepage centralizes navigation. They cannot establish whether these features improve achievement or engagement. No student grades, submissions, surveys, interviews, demographic records, or personally identifiable student information were analyzed.

### 4.2 Artifact corpus and scope

The corpus consists of 18 instructional artifacts from the Fall 2026 COS 107 environment: the syllabus, the tablet-inspired homepage, and Labs 1-16. We examined the syllabus for learning objectives, sequencing, assessment, AI policy, and Hallmark alignment; the homepage for navigation, multimedia, and links across environments; and the labs for their learning sequences, guided and independent tasks, debugging support, interactive extensions, testing, and technical progression.

The scope is deliberately limited. The syllabus mentions 15 optional challenges and additional Canvas-only materials, but their internal content was not included in the corpus. We describe their place in the course without attributing unobserved instructional features to them. The public course environment makes the reviewed design accessible, while protected Canvas content and student work are excluded from this analysis.

### 4.3 Analytic procedure

We analyzed the artifacts in three passes. First, we inventoried major sections, learning activities, technologies, assessments, and learner supports. Second, we organized evidence in a cross-artifact matrix using eight dimensions: orientation and coherence, problem representation, scaffolding, guided-to-independent transfer, interactive application development, debugging and testing, AI function and governance, and assessment integration. Third, we compared recurring patterns with departures from those patterns across the semester.

To make recurrence transparent, we completed a structural audit of Labs 1-15. We recorded whether each lab included (a) IPO problem representation, (b) a guided task, (c) an independent task, (d) a Colab/Gradio extension, (e) debugging or common-error guidance, (f) a task checklist, and (g) submission requirements. Lab 16 was examined separately because its public HTML structure differs from the established pattern.

The course designer was also involved in the analysis. This provided direct access to design rationale but also introduced potential positive-design bias. To mitigate this risk, we restricted descriptive findings to features observable in the artifacts or explicitly stated in the syllabus, reported structural counts where possible, and excluded claims of effectiveness. We also treated unresolved design tensions as part of the findings.

### 4.4 Semester task progression

Table 1 pairs the guided and independent activities across Labs 1-15. The second task is designed to require adaptation of a familiar computational structure to a different context rather than simple reproduction of the instructor example.

*Table 1. Guided and independent task progression across Labs 1-15.*

| Lab | Programming focus | Guided / in-class task | Independent transfer task |
|---|---|---|---|
| **1** | Print, comments, strings | Personal Greeting Card | Event Announcement |
| **2** | Input and variables | Student Profile Builder | Club Membership Form |
| **3** | Numerical calculations | Restaurant Tip Calculator | Sales Tax Calculator |
| **4** | Decisions | Grade Advisor | Movie Ticket Price |
| **5** | Loops | Weekly Savings Tracker | Study Time Tracker |
| **6** | Functions | Kilometer-to-Mile Converter | Celsius-to-Fahrenheit Converter |
| **7** | Modules and random numbers | Number Guessing Game | Dice Roller |
| **8** | Strings | Simple Password Checker | Email Username Extractor |
| **9** | Lists | Score Analyzer | Grocery Price Analyzer |
| **10** | Dictionaries and sets | Course Code Lookup | Mini Vocabulary Dictionary |
| **11** | Files and exceptions | Daily Journal Save | Expense Log |
| **12** | CSV data | Class Survey Summary | Simple Temperature CSV |
| **13** | Matplotlib visualization | Study Hours Bar Chart | Weekly Spending Chart |
| **14** | Classes and objects | Pet Class | Book Class |
| **15** | SQLite | Simple Task Tracker Database | Contact Database |

# 5. Findings

## 5.1 A stable course-level orientation layer

The tablet-inspired homepage gives different course functions recognizable locations while keeping Canvas as the system for official coursework. The browser-like area provides syllabus access; the music-style area contains audiovisual course communication; the mail-style area displays course points; and the settings-style list links to the Introduction and Weeks 1-16. These locations offer a consistent entry point even when programming activities take place in different tools.

The homepage is important for orientation, not for how closely it resembles a consumer tablet. Its purpose is to centralize course access while leaving specialized tools in place. However, decorative device cues may also introduce distraction. Whether the interface actually reduces navigation effort remains an empirical question.

## 5.2 A repeated Lab 1-15 learning grammar

The clearest cross-artifact pattern is the consistency of Labs 1-15. All 15 include IPO problem representation, instructor-guided programming, an independent task, a Colab/Gradio extension, debugging or common-error guidance, a checklist, and explicit submission requirements. Topics change, but the expected learning process remains recognizable.

The repeated sequence is straightforward: understand a concept, represent the problem, develop code with support, interpret errors, solve a related problem independently, build an interactive extension, test, and submit. This consistency stabilizes the process while technical demands increase from basic output to collections, files, visualization, classes, and databases.

## 5.3 Two forms of transfer are designed into the weekly sequence

Table 1 shows that transfer is incorporated into weekly practice rather than postponed until the final project. Each guided task is paired with an independent task that uses a similar computational structure in a different setting. For example, a restaurant-tip calculator becomes a sales-tax calculator, a grade advisor becomes a movie-ticket pricing program, and a Pet class becomes a Book class. The conceptual connection is visible, but students must adapt it.

The second transfer occurs through Gradio. Guided-to-independent work changes the problem context; a console-to-interface extension changes the form in which the solution is used. Students reorganize Python logic into a function and connect it to interface controls. Thus, the course repeatedly incorporates both problem-context transfer and interface-representation transfer.

By Lab 15, this progression reaches a SQLite task tracker with add, update, delete, and view operations. The cumulative message is that introductory constructs are not isolated syntax exercises; they can be combined into functional, interactive software.

### 5.4 Debugging and testing are treated as continuous problem-solving practices

Debugging is built into the lab sequence rather than taught only as a separate topic. Early tasks ask students to locate an error and compare expected with observed behavior. Later checklists ask whether the program executes, output is clear, variable names are meaningful, and at least two test cases have been tried. Common-error guidance becomes more specific as programming topics become more complex.

This approach treats errors as evidence for revising a solution. Students compare intended behavior with actual behavior and reconsider their implementation or mental model. Such repeated checking is consistent with research emphasizing metacognition and self-regulation in programming [5,6].

### 5.5 AI operates in two pedagogically distinct roles

The analysis distinguishes two roles for AI. Instructor-side AI supports course communication. The homepage includes a video area for course-structure materials, and the broader design draws on singing-syllabus and talking-avatar approaches developed by the instructor-designer (Wu, 2025, 2026). These formats are intended to supplement accessible written materials, not replace them.

Student-side AI assistance requires different safeguards because it can perform portions of the assessed work. The syllabus permits AI use for assignments only when explicitly authorized and requires detailed disclosure of any permitted use. Students must explain where, what, why, and how AI contributed. The policy names activities such as brainstorming, code assistance, debugging, data analysis, visualization, outlining, and editing as examples requiring careful disclosure when authorized.

This functional distinction changes the question from whether AI is present to what AI is doing and who remains responsible for the intellectual work. An instructor-generated video changes the format of a course message; a coding assistant can change who develops the solution. These cases therefore require different evaluation criteria and different governance rules.

### 5.6 Lab 16 functions as a deliberate structural break and scaffold-fading culmination

Lab 16 deliberately departs from the earlier lab structure. Rather than presenting separate guided, independent, Gradio, and checklist sections, the reviewed public page introduces a substantially larger Pygame-based Flappy Bird codebase. Bird, Pipe, and Ground classes coordinate state and behavior through an event loop, sprite groups, randomized pipe generation, movement, media assets, and collision detection.

We interpret this shift as an instance of scaffold fading: earlier labs divide programming into explicit stages, whereas the culminating artifact combines previously separated constructs in one system. This interpretation has an important boundary. Additional Hallmark directions may be available in Canvas, but the reviewed public Lab 16 page does not display the same checklist, independent-task, Gradio, and debugging sections as Labs 1-15.

The syllabus connects Lab 16 to Essential Skill #2 and the AAC&U Problem Solving VALUE Rubric. As a result, the culminating work is intended to assess more than whether a game runs. It creates an opportunity to examine how students define problems, plan changes, test behavior, justify decisions, and evaluate their solutions. Whether these activities are successfully demonstrated must be determined through student work in a future evaluation.

**Note 1:** Artificial Intelligence (AI) Tool Usage - If you use any AI tool at any point in your work, you must clearly and fully disclose its use in your assignment, report, project submission, or any other deliverable. Your disclosure must be specific and detailed, including 📢

- Where AI was used (e.g., summarizing readings, generating ideas, code assistance, debugging, data analysis, data visualization, outlining, editing).
- What tool(s) were used (tool name and, when available, version/model, plus platform/app).
- Why you chose to use the tool (e.g., brainstorming, translation, grammar checking, accelerating coding, improving clarity).
- How you used the output (e.g., modified/edited, cited/quoted, directly included, used only as a reference, used to compare against your own work).
- Other details as needed. 😊 The instructor reserves the right to impose additional disclosure requirements as necessary.

🚨 Important: Failure to disclose AI use—or presenting AI-generated work as entirely your own without clear clarification—may be considered a violation of academic integrity.

**Note 2: Responsible and Legal Use**

- Some software, tools, and programs used in this course may involve privacy, security, ethical, or legal risks if they are misused. Students are expected t use all course technologies responsibly, ethically, and in compliance with applicable laws, university policies, and course instructions.
- All materials, demonstrations, and activities are provided for educational purposes only. Do not use them to access systems without permission, harm others, violate privacy, disrupt services, or engage in any unauthorized or illegal activity. You are responsible for how you use the knowledge and tools introduced in this course.
- The materials provided on Canvas are intended for educational use only. Please do not share, upload, or distribute these materials on the internet or t others outside this course.

**Note 3**

- The instructor reserves the right to adjust the percentages/pts based on the number of assignments and quizzes, and other factors.

***Figure 2. AI-use policy artifact: disclosure is organized around where, what, why, and how AI contributed.***

## Hallmark Assignment - 10 Essential Skills

1. Communicate Effectively.
2. ✅ Think critically to solve problems and create new ideas and solutions.
3. Apply quantitative reasoning skills to analyze and solve numerical problems.
4. Interact effectively with people from different backgrounds.
5. Adapt to changing circumstances while leading and supporting others.
6. Perform professionally.
7. Engage in civic life to improve society.
8. Collaborate and work in teams.
9. Apply academic knowledge, skills, and abilities.
10. Use information for decision making.

---

Hallmark Assignment - Lab 16

- CPE Essential Skills #2 will be assessed using the **AACU VALUE Rubric: Problem Solving (Thoro 10)** Rubric.
- Thoro10 Essential Skills: 2. Think critically to solve problems and create new ideas and solutions.
  - Through this project, students will practice critical thinking by questioning assumptions, analyzing information, and making informed decisions. They will also apply creative thinking by combining ideas in new ways and designing original solutions to real-world challenges.

***Figure 3. Hallmark Assignment alignment in the syllabus: Lab 16 is linked to Essential Skill #2 and the AAC&U Problem Solving VALUE Rubric.***

## 6. Design Tensions and Tradeoffs

An informative design case must document tradeoffs as well as intended features. Table 2 summarizes the principal tensions evident in the artifacts, the design responses, and the evidence still needed to evaluate them.

***Table 2. Major design tensions, current responses, and unresolved evaluation needs.***

| Design tension | Current design response | Unresolved question / evidence needed |
|---|---|---|
| **Coherence vs. tool proliferation** | Tablet shell centralizes orientation; lab grammar remains stable across tools. | Does the shell reduce navigation effort, or does another interface add complexity? |
| **Authenticity vs. procedural overhead** | Gradio follows core Python logic rather than preceding it. | Do Colab/Gradio transitions strengthen transfer without overwhelming novices? |
| **Support vs. independence** | Guided examples are followed by related independent tasks; Lab 16 fades the weekly scaffold. | Is the transition to the Hallmark project appropriately calibrated? |
| **Manual typing vs. accessibility/efficiency** | Labs use no-copy code containers to encourage active typing. | Does this encourage attention, or create unnecessary barriers for students using accommodations or assistive technology? |
| **Multimodal engagement vs. distraction** | AI-assisted video/avatar media supplement written materials. | Which media improve communication, and which produce novelty or extraneous attention? |
| **AI support vs. cognitive outsourcing** | AI completion of assignments is restricted; authorized use requires where/what/why/how disclosure. | Can students evaluate AI output and distinguish support from substitution? |
| **Rubric alignment vs. activity-level visibility** | Syllabus clearly maps Lab 16 to AAC&U problem solving. | Should Hallmark criteria and reflection prompts be surfaced more explicitly on the Lab 16 page itself? |

## 7. Discussion

### 7.1 Progressive learner agency as the unifying mechanism

The principal contribution is the organization of the course around progressive learner agency rather than any single technology. Students begin with explicit problem representations, instructor modeling, debugging prompts, and closely related independent tasks. They then extend familiar program logic into interactive applications. Across the semester, technical demands increase while the learning process remains recognizable. Lab 16 removes many visible stages and introduces a larger integrated codebase.

This design shares the central concern of Use-Modify-Create and PRIMM: students should not be expected to move abruptly from explanation to open-ended creation. The study in [8] highlights the balance between structure and agency, while the argument in [9] moves students from prediction and investigation toward making. COS 107 applies a comparable progression across a semester and across several programming environments, without claiming to implement either framework directly.

### 7.2 Transfer is designed in two dimensions

The case makes two distinct transfer moves visible. The guided-to-independent pairing asks students to apply a computational pattern to a new problem. The Thonny-to-Gradio extension asks them to express similar logic through a different interface. The first changes the problem context; the second changes how a user interacts with the result.

Repeating both forms of transfer within a stable lab structure may allow students to focus on adaptation rather than procedural uncertainty. This is a design rationale, not a demonstrated effect. Future studies can test whether these two forms of transfer contribute differently to independent problem solving.

### 7.3 Navigation is pedagogical infrastructure, not decoration

The tablet-inspired shell is best understood as infrastructure for orientation. Programming courses often divide activity among a learning management system, an editor, a cloud notebook, and external resources. A course-specific gateway can organize these activities without trying to merge their technical functions. The broader principle is to centralize navigation while allowing specialized execution environments to remain separate.

A familiar interface metaphor does not itself demonstrate usability. It may help some students find resources while distracting others. The current analysis therefore describes the tablet shell as a proposed design response, not a proven intervention.

### 7.4 Responsible AI integration requires functional separation and disciplinary competence

The distinction between instructor-side and student-side AI is potentially transferable beyond programming. Course media generated with AI and an AI assistant producing a graded programming solution may use related technology, but they perform different educational functions. A course policy that treats every instance of AI as equivalent obscures this difference.

The syllabus addresses student assistance through authorization and disclosure. Authorization protects activities in which the student must perform the targeted reasoning. The where/what/why/how disclosure structure helps make permitted AI contributions visible. This approach is consistent with the argument in [13] that AI permissions should be aligned with the intellectual work being assessed.

Responsible AI use also depends on programming knowledge. Students need enough understanding of loops, functions, data structures, and program behavior to assess generated suggestions. This dependency helps explain why AI assistance can have different consequences for different novices [11]. In this context, AI literacy is closely connected to programming literacy.

### 7.5 The Hallmark Assignment exposes the boundary of the scaffold

The transition to Lab 16 is both useful and challenging. It provides an integrative endpoint instead of repeating a highly segmented task indefinitely. However, the public artifact reduces several supports at once. If detailed instructions and rubric guidance appear mainly in Canvas, students must coordinate not only the programming concepts but also directions distributed across platforms.

A future version could preserve the complexity of the Pygame project while making evidence of problem solving explicit. For example, students could identify a desired change, justify a strategy, implement and test it, document revisions, and evaluate the resulting behavior. Such prompts would make the AAC&U-aligned reasoning visible without supplying the solution.

## 8. Contributions and Implications for Practice

Four design ideas emerge from the case. First, course-level orientation is an instructional concern when multiple tools are necessary. Second, a repeated lab structure can make expectations clear while the programming content becomes more advanced. Third, transfer can be built into each lab both as independent application of a concept and as interface extension. Fourth, AI policies should distinguish communication from task assistance and make student responsibility explicit. These are design propositions supported by the artifact analysis, not measured effects.

The ideas are not tied to a particular software stack. Other learning management systems, Python editors, cloud environments, interface frameworks, or navigation designs could serve similar purposes. What matters is the relationship among the tools and the sequence of learning activities.

For instructors developing a similar course, the design suggests a practical sequence: organize access to resources, teach problem representation explicitly, model the reasoning process, require related independent practice, normalize debugging, connect core logic to an interactive artifact, define task-specific AI permissions, require disclosure when AI is authorized, and reduce scaffolding deliberately as students approach an integrative task.

## 9. Limitations and Future Research

This study has several limitations. First, it analyzes course artifacts rather than student outcomes. It cannot show that the ecosystem improves programming achievement, engagement, confidence, retention, or problem-solving performance relative to another approach. The structural counts describe what appears in the materials, not how students use those materials.

Second, the authors designed and analyzed the course. This gives direct access to design intent but creates a risk of confirmation bias. We limited claims to observable features, examined Lab 16 as a departure from the regular pattern, and made unresolved issues explicit. Independent coding or external design review would strengthen future analyses.

Third, the design reflects one course at one institution. Students' prior experience, class size, technology access, accessibility requirements, and institutional policies may differ elsewhere. The most transferable elements are therefore the design principles, not exact replication of the tools or activities.

Fourth, usability and accessibility have not been tested systematically. The tablet shell may simplify access or add distraction. Some lab code blocks restrict copying to encourage deliberate typing, but that choice may disadvantage students using assistive technology or accommodations. Animated and AI-generated media should likewise be supplemented with accessible text, captions, or transcripts and evaluated for instructional relevance.

Fifth, requiring AI disclosure does not guarantee appropriate AI use. Even a transparent submission may rely on AI for reasoning the assignment was intended to assess. Future studies should consider whether students can explain, test, modify, and defend AI-supported code—not only whether they report using AI.

Future empirical work should be planned prospectively and follow applicable institutional human-subjects procedures. A mixed-methods study could examine programming knowledge, debugging strategies, performance on lab tasks, Hallmark rubric evidence, and student perceptions of navigation, scaffolding, Gradio extensions, and multimodal communication. Where permitted, AI-use disclosures could be analyzed by purpose, such as explanation, debugging, verification, or code generation, rather than by frequency alone.

The Hallmark Assignment is a particularly useful opportunity for direct assessment. Future instructions could ask students to document problem definition, strategy selection, implementation, testing, evaluation, and reflection, making the problem-solving dimensions visible in their work. This would strengthen both instructional alignment and future empirical research.

## 10. Conclusion

Introductory programming continues to require careful attention to computational reasoning, testing, and debugging. Yet the environments in which students learn now include cloud notebooks, lightweight interfaces, multimedia, and GenAI systems capable of generating code. The instructional challenge is to use these tools without replacing the intellectual work through which beginners develop expertise.

COS 107 responds by organizing these technologies as a learning ecosystem. The tablet-inspired homepage provides an entry point. Labs 1-15 repeat a sequence of IPO representation, guided programming, debugging, independent transfer, Colab/Gradio extension, testing, and submission. Lab 16 then departs from that structure with a larger Pygame Hallmark project aligned with problem-solving assessment.

The key design idea is progressive learner agency. Students are expected to move from interpreting examples to adapting, testing, and integrating solutions. The course also distinguishes AI used to communicate about learning from AI used to complete learning tasks. This distinction supports more precise decisions about authorization, disclosure, and responsibility.

The aim is not to maximize the number of technologies in a course. It is to organize them around meaningful learning activities. The present artifact analysis offers a reusable design precedent, while leaving the question of effectiveness open for future empirical study. In introductory programming, technologies should help students explain, revise, and evaluate computational solutions—not eliminate the reasoning they are learning to perform.

## Declarations

### Ethics and human-subjects statement

This paper examines publicly shareable course-design artifacts. It does not use student grades, submissions, surveys, interviews, demographic records, or personally identifiable student information. Any future study of student performance or experience should follow applicable institutional human-subjects review or determination before research data are collected or analyzed.

### Materials availability

The public COS 107 learning environment and weekly materials are available at https://xinxingwu-uk.github.io/fall2026/COS107/Homepage/index.html. Institutional Canvas materials and protected student information were not part of the analyzed corpus.

### Generative AI assistance disclosure

OpenAI ChatGPT was used during manuscript preparation for language editing and Word-document refinement. The authors supplied the course artifacts and underlying design decisions. No student grades, submissions, or personally identifiable student information were provided to the AI system.